\documentclass{iopart}
\usepackage{graphicx}
\begin{document}

\title[Radiofrequency multipole traps]{Radiofrequency multipole traps: Tools
for spectroscopy and dynamics of cold molecular ions}

\author{Roland Wester}

\address{Department of Physics, University of Freiburg, Hermann-Herder-Str.\
 3, 79104 Freiburg, Germany}
\ead{roland.wester@physik.uni-freiburg.de}

\begin{abstract}
  Multipole radiofrequency ion traps are a highly versatile tool to study
  molecular ions and their interactions in a well-controllable environment. In
  particular the cryogenic 22-pole ion trap configuration is used to study
  ion-molecule reactions and complex molecular spectroscopy at temperatures
  between few Kelvin and room temperatures. This article presents a tutorial
  on radiofrequency ion trapping in multipole electrode configurations. Stable
  trapping conditions and buffer gas cooling, as well as important heating
  mechanisms, are discussed. In addition, selected experimental studies on
  cation and anion-molecule reactions and on spectroscopy of trapped ions are
  reviewed. Starting from these studies an outlook on the future of multipole
  ion trap research is given.
\end{abstract}

\submitto{\JPB}
\maketitle


\section{Introduction}

Advances in our understanding of the structure and dynamics of small quantum
systems is closely related to improved techniques to prepare such systems in
well-controlled quantum states. Particle traps that confine atoms, molecules
or clusters for a long time are essential tools for this. Traps for charged
particles, in particular the radiofrequency Paul trap
\cite{paul1958,paul1990:rmp} and the electromagnetic Penning trap
\cite{dehmelt1967:adv}, have been introduced many years ago. Two newer trap
designs have been developed recently. The electrostatic resonator Zajfman trap
has been introduced as a tool to study interactions of trapped ions in motion
\cite{zajfman1997:pra} and the Orbitrap has been developed as a versatile
high-resolution mass spectrometer \cite{hu2005:jms}. Ion traps find many
applications, among them mass spectrometry
\cite{blaum2006:pr,douglas2005:msr}, precision spectroscopy
\cite{margolis2004:sci,rosenband2008:sci}, and the implementation of quantum
logic \cite{leibfried2003:rmp}. Multipole radiofrequency ion traps, the topic
of this article, are widely used to control trapped molecular systems that
have many degrees of freedom \cite{gerlich1992:adv}. The spectroscopy and
reactivity of molecular ions is studied in these traps to unravel their
dynamics \cite{asvany2005:sci,otto2008:prl} and simulate the properties of
cold plasmas, such as the Earth's atmosphere or the interstellar medium
\cite{gerlich2006:ps}.

Radiofrequency ions traps have effective trap depths of the order of 1\,eV, in
contrast to traps for neutral atoms or molecules, which are based on
interactions of the electric or magnetic dipole moment or the electric
polarizability and are typically of the order of 10$^{-4}$ to 10$^{-7}$\,eV
\cite{grimm2000:adv}. This implies that ions that carry a significant amount
of kinetic energy can still be trapped. Different approaches exist to cool
trapped ions, collisional cooling with a surrounding neutral buffer gas
\cite{dehmelt1967:adv}, laser-cooling on a suitable optical transition
\cite{neuhauser1978:prl}, and - as a combination of these two - sympathetic
cooling with jointly trapped laser-cooled ions \cite{roth2008:arxiv}. Direct
laser cooling of molecules with their manifold of internal quantum states is
not possible due to the lack of a closed optical cycling transition. Therefore
either the first or the third method is employed to cool {\em molecular ions}.

The technique of sympathetic cooling of molecular ions with laser-cooled
atomic ions in a joint quadrupole ion trap has been introduced only a few
years ago \cite{molhave2000:pra,drewsen2004:prl,blythe2005:prl}. It has been
successfully used to cool the translational motion of trapped molecular ions
down to temperatures of a few millikelvin. These extremely cold and localized
ions are well suited for photodissociation experiments
\cite{bertelsen2004:epd,hojbjerre2008:pra} and for precision spectroscopy
\cite{koelemeij2007:prl}. However, ion-ion collisions have been found to leave
the internal quantum state distribution unchanged
\cite{bertelsen2006:jpb,koelemeij2007:pra} and are thus not useful for
rovibrational cooling of trapped molecular ions. As an alternative, an optical
pumping scheme has been proposed to cool internal degrees of freedom
\cite{vogelius2002:prl}. Recently, first reaction experiments with
sympathetically cooled molecular ions have been reported using room
temperature neutral reactants \cite{staanum2008:prl,roth2008:pra}, and - in
combination with electrostatically filtered molecules - at relative energies
corresponding to about one Kelvin \cite{willitsch2008:prl,willitsch2008:pccp}.

In the main part of this article the focus will be on collisional cooling with
a cold neutral buffer gas. In contrast to sympathetic ion-ion cooling this
approach allows one to cool the translational motion as well as the internal
rovibrational degrees of freedom at the same time. With helium as the buffer
gas, translational and internal temperatures down to a few Kelvin are reached
(see section \ref{buffergas:sect}). Details of the interaction of the
molecular ion with the helium, such as inelastic collision cross sections,
determine the time scale for buffer gas cooling but do not inhibit the
application in principle. This technique is therefore applicable to
essentially all stable molecular ions over a wide range of temperatures. In
addition, neutral collision partners for chemical reactions are also
thermalized to the same temperature. Buffer gas cooling therefore provides
good starting conditions for a variety of experiments on ion-neutral reactions
and laser spectroscopy under controlled thermal conditions (see sections
\ref{reactions:sect} and \ref{spectroscopy:sect}).

As explained in section \ref{buffergas:sect}, buffer gas cooling works best if
the interaction of the ion with the confining electric field of the
radiofrequency trap is limited to a minimum. This obvious contradiction is
resolved by trapping in higher-order radiofrequency multipole
configurations. These offer a suppressed electric field in the center of the
trap and an effective potential that increases much more steeply than the
harmonic potential of a quadrupole or Paul ion trap. In this way the
confinement in a multipole ion trap comes close to the particle-in-a-box
situation with ions moving in an almost field-free volume bound by steep
potential walls.

The application of higher order radiofrequency multipole structures for the
confinement of charged particles dates back to the 1970ies when the first
experiments with radiofrequency multipole ion guides were reported
\cite{teloy74}. The two-dimensional confinement in ion guides allowed precise
studies of ion-molecule collision cross sections at relative energies between
a few meV and several eV. This has developed into a widely used method up to
the present day \cite{armentrout2001:annu}. Three-dimensional confinement in
multipole structures has been introduced about a decade later. In octupole ion
traps experiments with laser-cooled ions have been carried out to image the
ion density by spatially resolving the fluorescence signal
\cite{schubert1989:josa,walz94}. More recently, novel ring-shaped and
cylindrical crystals made of laser-cooled ions have been studied in an
octupole ion trap \cite{okada2007:pra}. The first ion trap with a large
field-free region that was employed in experiments was the ring electrode trap
\cite{gerlich1989:apj}, which could already be cooled to liquid nitrogen
temperature. This lead to extended interaction times of the trapped ions of
many seconds instead of just microseconds. Thereby processes that occur with
very small cross sections could be studied. With the development of improved
cryogenic ion traps, in particular the 22-pole ion trap \cite{gerlich1995:ps},
processes such as radiative association \cite{gerlich1992:cr} and three-body,
complex-forming collisions \cite{paul1995:ijm,paul1996:cp} could be studied at
low temperature. This also made it possible to investigate low-temperature
reactions that are relevant for the understanding of the interstellar
medium. In the last ten years the 22-pole ion trap has become increasingly
popular to study ion-molecule reactions and molecular ion spectroscopy.

The concepts of ion guiding and trapping with multipole fields have been
discussed extensively in a landmark review by Gerlich in 1992
\cite{gerlich1992:adv}. Since then major developments have happened, most
notably the introduction of the widely used 22-pole ion trap
\cite{gerlich1995:ps}. More recently, a few overview articles have reviewed
multipole ion traps \cite{gerlich2003:hi,gerlich2008:book} and their
application in mass spectrometry \cite{douglas2005:msr} and astrophysics
\cite{gerlich2002:pss,gerlich2006:ps}.

This tutorial on multipole rf ion traps starts out presenting the concept of
trapping in high-order multipole fields (next section), followed by a
discussion of ion cooling with buffer gas (section \ref{buffergas:sect}). Then
the typically employed experimental techniques are described (section
\ref{experimental:sect}). The next two sections are devoted to the two major
research areas in which multipole ion traps are used, cold ionic reactions
(section \ref{reactions:sect}) and spectroscopy of internally cold ions
(section \ref{spectroscopy:sect}). The outlook (section \ref{outlook:sect})
highlights several interesting directions that are currently coming into the
focus of this research.


\section{Trapping ions in high-order multipole fields \label{multipole:sec}}

\subsection{The effective potential approximation}

\begin{figure}[tb]
\begin{center}
\includegraphics[width=\columnwidth]{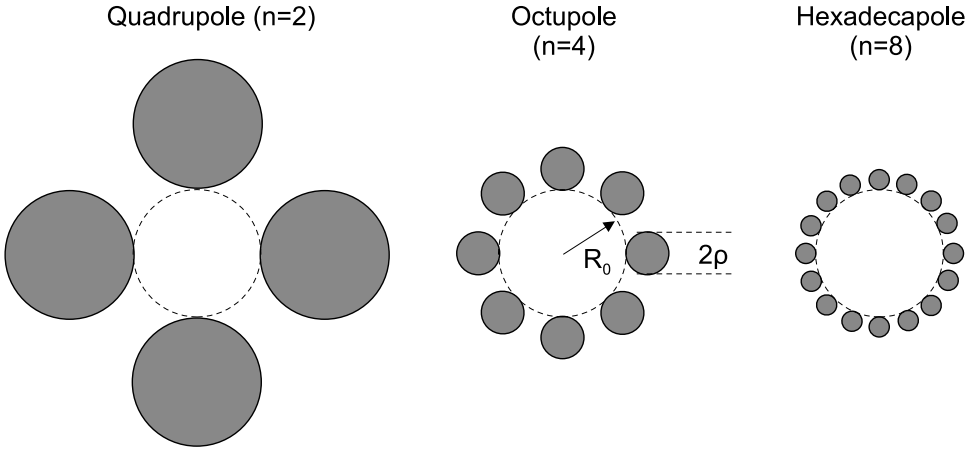}
\caption{\label{fig-multipoles} Cross sections through three different linear
  radiofrequency multipole structures of order $n=2$, 4 and 8. All multipoles
  are approximated by $2n$ cylindrical rods of optimal diameter (see Eq.\
  \ref{optimal-diameter:eq}).}
\end{center}
\end{figure}

Multipole ion traps and ion guides typically employ linear radiofrequency
electrode geometries that follow a discrete cylindrical symmetry. The
time-varying electric potential of an ideal two-dimensional multipole
configuration of order $n$ with infinite extension along the $z$-direction is
given in cylindrical coordinates by
\begin{equation}
V(r,\phi,t) = V_0 \cos(n \phi) \left(\frac{r}{R_0}\right)^n \sin (\omega t),
\label{multipole:eq}
\end{equation}
where $V_0$ is the radiofrequency amplitude, $R_0$ the inscribed radius of the
rf electrodes (see Fig.\ \ref{fig-multipoles}) and $\omega$ denotes the
angular frequency. In practice, this potential is created by $2n$ cylindrical
electrodes of radius $\rho$ arranged on a cylinder of inscribed radius $R_0$,
as shown in Fig.\ \ref{fig-multipoles} for several different multipole orders
$n$. To achieve a potential that optimally approximates the curvature of the
ideal multipole potential of Eq.\ \ref{multipole:eq} to first and second
order, the diameter of the rods has to fulfill the relation
\cite{gerlich1992:adv}
\begin{equation}
\rho = \frac{R_0}{n-1}.
\label{optimal-diameter:eq}
\end{equation}
For a trap diameter of $2 R_0 = 10$\,mm and a rod diameter of $2\rho = 1$\,mm
this leads to $n=11$, the widely-used 22-pole ion trap.

The dynamics of ions in the time-dependent field of a multipole ion trap can
not be solved analytically, in contrast to the dynamics in a quadrupole or
Paul trap, because the equation of motion becomes non-linear in the position
coordinate. Instead, the best way to describe the dynamics is with the aid of
the effective potential approximation, which was originally introduced for rf
traps by Dehmelt \cite{dehmelt1967:adv}. For this one assumes that the
position coordinate of a trapped ion is composed of a rapid oscillation $\vec
\xi(t)$ that follows the radiofrequency, the micromotion, and a slowly varying
drift motion $\vec R(t)$, also referred to as the secular motion,
\begin{equation}
\vec r(t) = \vec R(t) + \vec \xi(t).
\end{equation}
The electric field acting on the ion is expanded in a Taylor series, which
leads to the equation of motion
\begin{equation}
m \frac{d^2}{dt^2} \vec R 
+ 
m \frac{d^2}{dt^2} \vec \xi 
= 
q \vec E(\vec R,t)
+ 
q (\vec \xi \vec \nabla) \vec E(\vec R,t).
\label{motion:eq}
\end{equation}
For short times, of the order of the rf oscillation period, the first and the
fourth term of Eq.\ \ref{motion:eq} may be neglected. This results in a
differential equation for $\vec \xi(t)$ that is solved by $\vec \xi(t) = -\vec
\xi_0 sin(\omega t)$ with $\vec \xi_0 = \frac{q \vec E(\vec R)}{m
\omega^2}$. This describes the micromotion of an ion in the rf field. By
averaging Eq.\ \ref{motion:eq} over one rf oscillation period and using the
solution for $\vec \xi$, the second and third terms vanish. This leads to an
equation of motion for the drift coordinate $\vec R$
\begin{equation}
m \frac{d^2}{dt^2} \vec R = \vec \nabla V_{\rm eff}(\vec R),
\end{equation}
with an effective conservative potential
\begin{equation}
V_{\rm eff}(\vec R) 
= 
\frac{q^2}{2 m \omega^2} \langle \vec E(\vec R,t)^2 \rangle.
\end{equation}
The angle brackets denote the time average over one oscillation period.

For the ideal two-dimensional multipole field of Eq.\ \ref {multipole:eq} one
obtains the simple, cylindrically symmetric expression for the effective
potential
\begin{equation}
V_{\rm eff}(\vec R) 
= 
\frac{q^2 n^2 V_0^2}{4 m \omega^2 R_0^2} \left(\frac{R}{R_0}\right)^{(2n-2)}.
\label{eff_potential:eq}
\end{equation}
For large $n$ this potential increases rapidly with the increasing radius $R$
and approximates an effective potential with a large almost field-free central
region and steep walls.

The effective potential is intimately related to the ion's micromotion. In
fact, from the solution for $\vec \xi(t)$ it follows that the time-averaged
kinetic energy of the micromotion equals exactly the effective potential
\begin{equation}
V_{\rm eff}(\vec R) = \langle \frac{1}{2} m \dot{\vec \xi}^2 \rangle.
\end{equation}
This shows that within the effective potential approximation confinement in
the ion trap occurs, because kinetic energy of the drift motion is reversibly
converted into kinetic energy of the micromotion, leading to a maximum radius
$R_{\rm max}$ of the ion's drift motion.

\subsection{Stable operating conditions}

The effective potential description breaks down when the amplitude of the
micromotion becomes too large for the truncated Taylor expansion to be a valid
approximation. In this case energy exchange between the micromotion and the
secular motion occurs that increases the amplitude of the secular motion until
the ion is lost in a collision with the trap electrodes. This coupling is
important when the electric field amplitude changes significantly over the
distance traveled by an ion during one cycle of the micromotion, i.e.\ when
the stability parameter, introduced by Teloy and Gerlich
\cite{teloy74,gerlich1992:adv},
\begin{equation}
\eta = \frac{| 2 (\vec \xi_0 \vec \nabla) \vec E(\vec R )|}{| \vec E(\vec R )| }
     = \frac{2 |q|}{m \omega^2} \left| \vec \nabla | \vec E(\vec R) | \right|
\end{equation}
is no longer a small quantity. For an ideal multipole one obtains
\begin{equation}
\eta = 2 n (n-1) \frac{|q| V_0}{m \omega^2 R_0^2} \left(\frac{R}{R_0}\right)^{n-2}.
\end{equation}
Note that $\eta$ depends explicitly on the radius of the secular motion for $n
> 2$. Only for a quadrupole ion trap one obtains a constant value, which in
this case equals the parameter $q$ that enters into the Mathieu differential
equation \cite{paul1958,leibfried2003:rmp}. The solutions of the Mathieu
equations yield the $q$-intervals for stable and unstable ion trapping. In
multipole ion traps a clear distinction between stable and unstable regimes of
trap operation is not possible \cite{haegg86a,haegg86b}. Instead numerical
simulations have been used to define stable operating conditions: $\eta < 0.3$
for $R/R_0 < 0.8$ \cite{gerlich1986:proc,gerlich1992:adv}.

\begin{figure}[tb]
\begin{center}
\includegraphics[width=0.7\columnwidth]{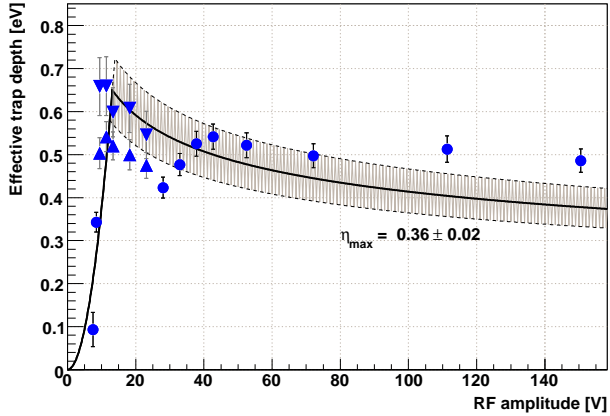}
\caption{\label{fig-evap} (Color online) Experimentally deduced effective
  depth of a 22-pole ion trap for Cl$^-$ anions as a function of the rf
  amplitude. Each data point originates from a Boltzmann fit to
  temperature-dependent loss rates. The breakoff of the effective trap depth’s
  quadratic increase is attributed to the appearance of nonadiabatic ion
  motion in the trap. The solid line is a fit of a simple analytical model of
  the switch between adiabatic and nonadiabatic motion with the maximum
  stability parameter $\eta_{\rm max}$ as the only free parameter
  \cite{mikosch2008:pra}.}
\end{center}
\end{figure}

An experimental test of the stability condition in a multipole ion trap has
been carried out by studying the loss rate of ions from a 22-pole ion trap
\cite{mikosch2007:prl,mikosch2008:pra}. The loss rate was found to depend
exponentially on an activation energy that was associated with an effective
depth of the trapping potential. The measured effective trap depth as a
function of the rf amplitude is shown in Fig.\ \ref{fig-evap}. It follows Eq.\
\ref{eff_potential:eq} with $R=R_0$ for low amplitudes (with no free
parameter). Above 13\,V the trap depth does not increase further, which
indicates that for large values of $R$ $\eta$ is too large to provide stable
trapping. Numerical simulations \cite{mikosch2008:pra} show that if an ion
finds itself at positions in the trap with such a large $\eta$-value, the
effective potential approximation breaks down and the secular kinetic energy
of trapped ions is increased in the radiofrequency field. This {\em
instability heating} then leads to trap loss.

A simple evaporation model is used to describe the leveling-off of the
measured effective potential. It assumes a sudden change from stable to
unstable motion at a critical radius where $\eta$ becomes larger than a
critical value. This value is determined from a fit to the data to be
$0.36\pm0.02$ \cite{mikosch2008:pra}, in agreement with the safe operating
conditions defined above. The model for the effective trap depth strongly
simplifies the complex non-linear dynamics. Nevertheless, it provides direct
experimental evidence for the range of applicability of the effective
potential approximation in multipole ion traps.


\section{Preparing cold ions \label{buffergas:sect}}

\subsection{Buffer gas cooling}

The trapped ions are cooled in collisions with an inert buffer gas at a
well-controlled temperature. Typically helium is chosen as buffer gas, because
of its high vapor pressure at low temperature. The buffer gas equilibrates
with the gas inlet tubes and the copper housing of the ion trap to its
temperature. 

The standard model to estimate collision rates of ions with neutral atoms or
molecules is the Langevin or capture model \cite{levine2005}. The assumption
of this model is that a collision occurs with 100\% probability when the two
collision partners come closer to each other than a critical distance. The
impact parameter that belongs to this critical distance is defined by the
requirement that the relative kinetic energy is just sufficient to surmount
the centrifugal barrier at this impact parameter. For larger impact parameters
the ion-neutral interaction is neglected. This simplified picture yields the
Langevin cross section (in SI units)
\begin{equation} 
\sigma = \frac{|q|}{2 \epsilon_0} \sqrt{\frac{\alpha}{\mu v^2}},
\end{equation}
with the neutral polarizability $\alpha$, the reduced mass of the two-body
system $\mu$ and the relative velocity $v$ ($q$ is the ion charge and
$\epsilon_0$ the electric constant). In a thermal ensemble the collision rate
coefficient $k = \langle \sigma v \rangle$, obtained by averaging over the
velocity distribution, is the more relevant quantity. In the Langevin model
one obtains the velocity and temperature independent rate coefficient
\begin{equation}
k = \frac{|q|}{2 \epsilon_0} \sqrt{\frac{\alpha}{\mu}}.
\end{equation}

An ion trapped in a buffer gas of given temperature will thermalize within a
number of elastic collisions with the buffer gas. At a constant buffer gas
density $n_{\rm bg}$ the average time between successive collisions is given
by the inverse of the collision rate $(k n_{\rm bg})^{-1}$. Typical values lie
in the microsecond regime, given the typical values for the Langevin rate of
$10^{-9}$\,cm$^3$/s and for the density of $10^{12}$ to $10^{15}$\,cm$^{-3}$.
The number of elastic collisions that are required for the ion velocity
distribution to resemble a Maxwell-Boltzmann distribution is estimated with
numerical simulations. A Monte Carlo simulation is used to calculate ion
trajectories interacting with a buffer gas in free space. To simplify the
problem isotropic elastic scattering collisions are assumed. Such simulations
show that between three and ten collisions already lead to a distribution of
the ion velocity that closely resembles a Maxwell-Boltzmann distribution. Ions
that are trapped in a conservative, time-independent trapping potential also
acquire a thermal velocity distribution, described by the temperature of the
buffer gas, within only few buffer gas collisions.

\begin{figure}[tb]
\begin{center}
\includegraphics[width=0.7\columnwidth]{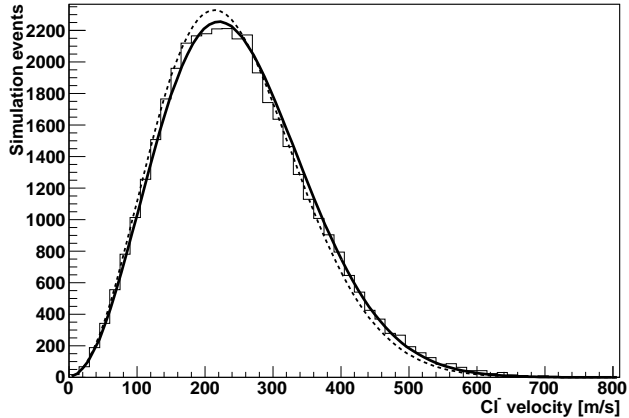}
\caption{\label{fig-simulation} Simulated velocity distribution of Cl$^-$
  anions in a 22-pole ion trap filled with 100\,K helium buffer gas. The
  distribution is fitted to a Maxwell-Boltzmann distribution with a
  translational temperature of $106\pm1$\,K (solid line), hotter than the
  buffer gas temperature due to radiofrequency heating. For comparison also
  the Maxwell-Boltzmann distribution for 100\,K is shown (dashed line).}
\end{center}
\end{figure}

\subsection{Radiofrequency heating}

In the true time-dependent electric field of a radiofrequency ion trap the
fast oscillatory micromotion perturbs the velocity distribution of trapped
ions and hinders them to reach the temperature of the buffer gas. This effect
of {\em radiofrequency heating} leads to higher translational temperatures of
the trapped ions than the buffer gas temperature. In spectroscopic studies of
trapped ions in 22-pole traps this effect has become visible
\cite{mikosch2004:jcp,kreckel2008:jcp}. Its dependence on different parameters
of the trap has recently been analyzed in detail using numerical simulations
\cite{asvany2009:ijm}.

We have calculated the effect of radiofrequency heating with a Monte Carlo
simulation that includes the time-dependent potential of an ideal multipole
(Eq.\ \ref{multipole:eq}). The simulation has been carried out for Cl$^-$ in
an infinitely long 22-pole ion trap ($n=11$), filled with 100\,K helium buffer
gas. The velocity distribution of trapped Cl$^-$ is extracted by following the
trajectory of a trapped ion and sampling its instantaneous velocity every two
to three collisions with the buffer gas. This samples the thermal ion velocity
distribution in the trap. The histogram in Fig.\ \ref{fig-simulation} shows
the obtained velocity distribution for a buffer gas collision rate of
10$^5$\,s$^{-1}$ and an rf amplitude $V_0 = 50$\,V. The histogram is fitted
with a Maxwell-Boltzmann distribution (solid line in Fig.\
\ref{fig-simulation}) where the ions' translational temperature is the only
free parameter. The numerical data agrees well with a Maxwell-Boltzmann
distribution, but the obtained temperature is $106\pm1$\,K, about 6\% hotter
than the temperature of the helium buffer gas. For comparison, the dashed line
shows a Maxwell-Boltzmann distribution for 100\,K.

To test the numerical accuracy of the Monte-Carlo simulation we have
calculated the velocity distribution for ions in the time-independent
effective potential of a trap with the same trapping parameters. Here a
temperature of $101\,\pm 1$\,K is obtained, in agreement with the expectation
of thermalization. The about 6\% increase of the temperature that is found for
the time-dependent radiofrequency field can therefore be attributed to
radiofrequency heating. It is caused by elastic collisions with buffer gas
atoms that interrupt the micromotion and thereby transfer energy from this
motion into the secular motion. Therefore the temperature mismatch decreases
with an increased ratio of the ion mass to the buffer gas mass. The
simulations also show that the relative temperature increase is independent of
the amplitude of the radiofrequency field, the temperature and the collision
rate of the buffer gas across the experimentally relevant parameter range.

The radiofrequency heating and therefore the increase of the effective ion
temperature compared to the buffer gas temperature changes notably with the
multipole configuration of the ion trap. The larger the multipole order $n$
the smaller is the effective heating rate. This is qualitatively explained by
the shape of the confining effective potential, which represents the kinetic
energy of the micromotion. The steeper the effective potential the more time
an ion spends in the almost field-free central region of the trap where buffer
gas collisions may not transfer micromotion energy into the secular
motion. For large $n$, such as in the 22-pole ion trap, radiofrequency heating
can occur only near the steep wall of the effective potential.

\subsection{Thermalization of internal degrees of freedom}

For trapped molecular and cluster ions the thermalization of the translational
motion to a temperature near the buffer gas temperature is only the first
step. Such complex ions also possess $3N-6$ ($3N-5$ for linear molecules )
internal vibrational degrees of freedom and three (two for linear molecules)
rotational degrees of freedom. Inelastic collisions with the buffer gas can be
employed to also thermalize these degrees of freedom.

For internal cooling, the relative kinetic energy of the ion-neutral collision
system is the relevant quantity that determines the ion's internal excitation
in steady state. In the case of weak radiofrequency heating the ion's
translational velocity distribution follows a Maxwell-Boltzmann distribution,
as discussed above. In this case the effective temperature of the relative
motion is given by
\begin{equation}
T_{\rm eff} = (m_n T_n + m_i T_i)/(m_n + m_i),
\end{equation}
where $T_n$ and $T_i$ are the translational temperature and $m_n$ and $m_i$
the masses of the neutral and the ion, respectively. This shows that heavy
ions and clusters in helium buffer gas may reach internal temperatures near
the buffer gas temperature, even if their translational temperature is
affected by radiofrequency heating. Experimentally, both rotational
\cite{Schlemmer1999Laser,dhzonson2006:jms} and vibrational cooling
\cite{boyarkin2006:jacs} have been observed.

The time scale for equilibration of the internal degrees of freedom is
typically estimated to be longer than for elastic collisions, because the
inelastic rate coefficient is often significantly smaller than the Langevin
limit. Precise values are rare for most molecular ions, but estimates range
between 10 to 10$^4$ collisions that are needed to achieve collisional
thermalization. Besides this mechanism, spontaneous radiative emission may
also remove internal energy from the trapped ions. This holds in particular
for stiff vibrational degrees of freedom where the radiative lifetimes lie in
the millisecond range \cite{amitay1998:sci,hechtfischer1998:prl}, which leads
to radiative cooling rates that are comparable with collisional cooling rates.


\section{Experimental techniques \label{experimental:sect}} 

\begin{figure}[tb]
\begin{center}
\includegraphics[width=0.7\columnwidth]{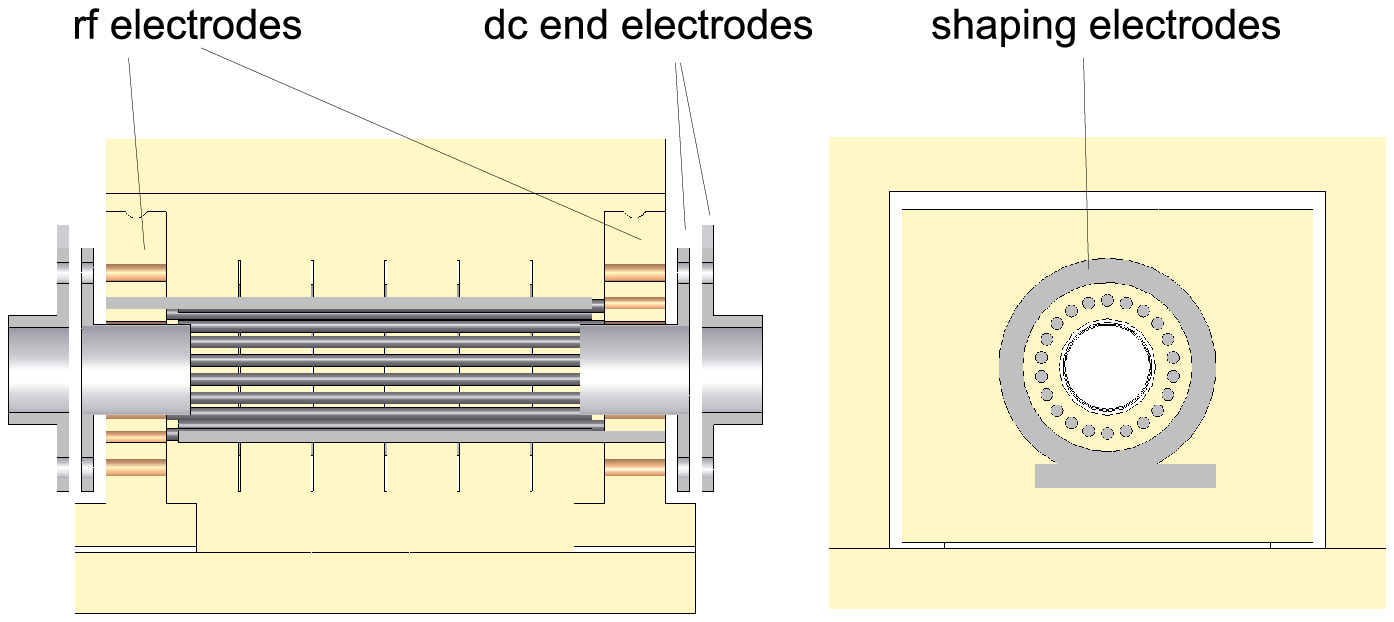}
\includegraphics[width=0.29\columnwidth]{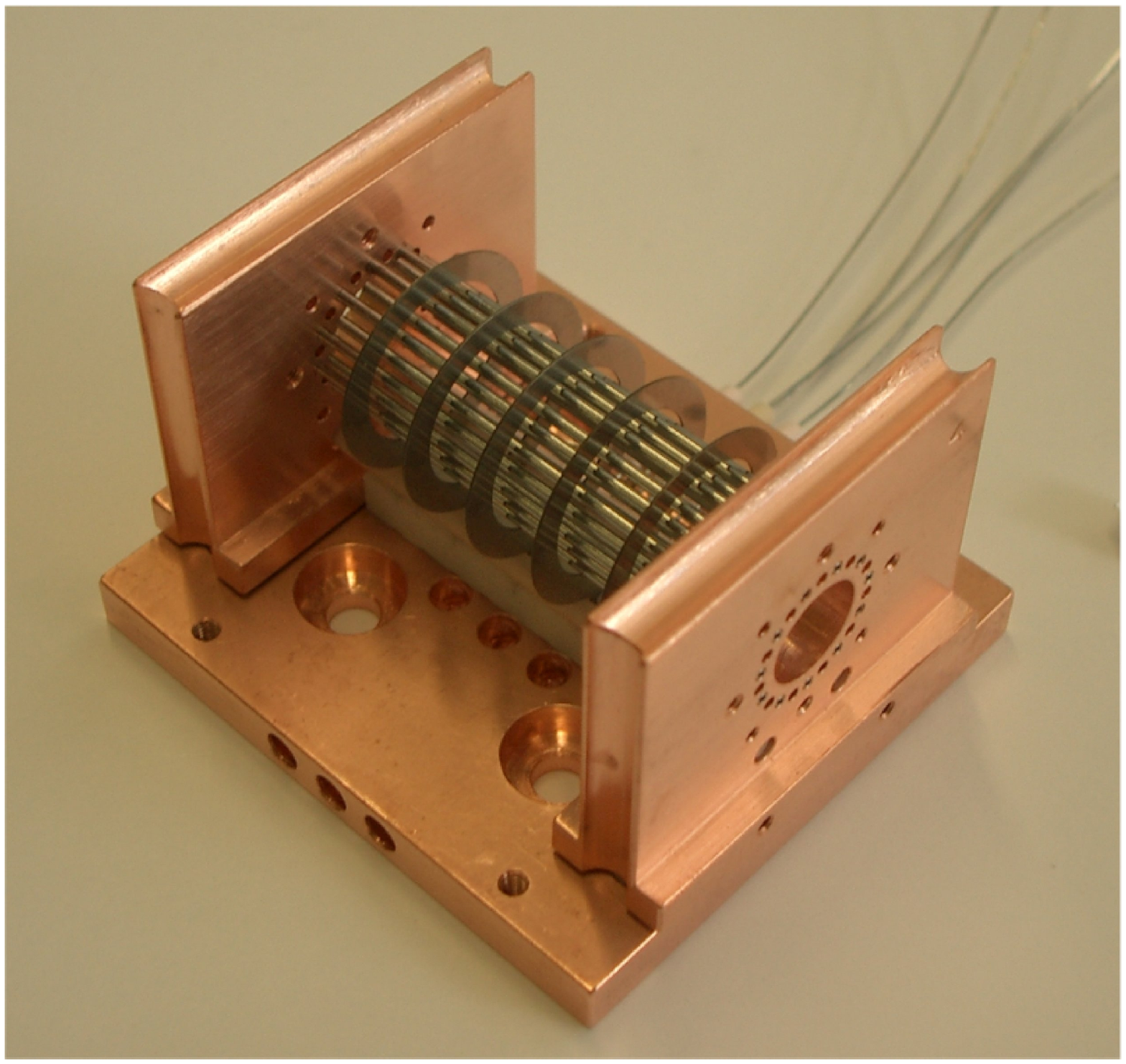}
\caption{\label{fig-22pole} (Color online) Cross section through the center of
  a 22-pole ion trap \cite{gerlich1995:ps}, viewed parallel to the symmetry
  axis (left panel) and perpendicular to it (middle panel). On the right a
  photograph of the Freiburg 22-pole ion trap is shown. The 22 rf electrodes,
  mounted alternatingly into the two side plates and two pairs of dc end
  electrodes provide ion confinement. In addition, five shaping electrodes are
  placed around the 22 poles. These electrodes are used to form the trapping
  potential along its symmetry axis and lead to temporally shorter ion packets
  after extraction from the trap \cite{mikosch2008:pra}.}
\end{center}
\end{figure}

The most widely used multipole ion trap configuration consists of $2n$
cylindrical stainless steel electrodes that are arranged around a cylinder
with inscribed radius $R_0$ (see Fig.\ \ref{fig-multipoles}). The employed
types of traps range from hexapoles and octupoles to 22-pole structures. A
photograph of the 22-pole trap \cite{gerlich1995:ps} is depicted in Fig.\
\ref{fig-22pole}. It is about 40\,mm in length with an inscribed radius of
5\,mm. Confinement in the direction parallel to the electrodes is provided by
two cylindrical electrostatic electrodes at the two ends of the trap.

A mechanical enclosure, typically made from copper, houses the ion trap and is
filled with buffer gas for collisional cooling of the trapped ions. The trap
wall temperature, which sets the buffer gas temperature, is changed by
mounting the entire trap on a cryostat. Most widely used are closed-cycle
helium cryostats of the Gifford McMahon type, which provide cooling powers of
the order of one watt near 10\,K. They are complemented with variable power
resistive heaters to increase the temperature to higher values in a controlled
way.

Ions are typically not formed in the cryogenic, high vacuum environment of the
trap itself. Instead they are created in a separate ion source and loaded into
the trap through one of the end electrodes. A variety of ion sources are used,
continuous electron impact ion sources, plasma sources, radiofrequency storage
ion sources, pulsed supersonic beam sources and electrospray ionization
sources. Typically the ions are created with less than 1\,eV kinetic energy,
accelerated gently to few eV, transported through a radiofrequency quadrupole
for mass selection, and then passed into the ion trap
\cite{gerlich1992:cr}. In an alternative approach, which we have developed to
improve flexibility, the ions are accelerated and mass-selected by
time-of-flight before the selected ion packet is decelerated and loaded into
the ion trap \cite{mikosch2008:pra}. Trapping occurs either by opening the
electrostatic end electrode for the entering ions and closing it after the
ions have entered the ion trap. Alternatively the ions are kept at a high
enough kinetic energy to overcome the potential of the static end electrode
and are then decelerated rapidly by buffer gas collisions inside the trap
before they have a chance to leave the trap again. The advantage of the latter
scheme is that it allows continuous loading of the ion trap over an extended
period of time. This is particularly advantageous when only weak currents of
the ion of interest are produced in the ion source.  

About 10$^2$ to 10$^4$ trapped ions are typically used in experiments. In this
case the Coulomb interaction between the trapped ions is not significant,
since the volume in which the ions are confined amounts to several
cm$^3$. Furthermore, under these conditions the charging of insulator surfaces
and saturation effects in the ion detector typically do not become important.

After a selected storage time, the extraction of the trapped ion ensemble
proceeds by lowering the potential of one of the electrostatic end electrodes.
To mass-analyze the composition of the extracted ions, either a second
quadrupole mass spectrometer \cite{gerlich1992:cr} or a second time-of-flight
mass spectrometer is employed \cite{mikosch2008:pra}. The ions are then
detected with a single-particle detector, either a Daly detector
\cite{daly60}, a channeltron detector or a microchannel plate
\cite{wiza1979:nim}. In all cases between about 50\% and more than 90\%
detection efficiency for single ions is obtained.


\section{Reaction dynamics of cold ions \label{reactions:sect}}

Many ion-molecule reactions have been studied in multipole ion traps (see
e.g.\ Ref.\
\cite{gerlich1989:apj,gerlich1992:cr,gerlich1995:ps,gerlich2002:pss,gerlich2006:ps}).
In these experiments buffer gas cooling of the translational and internal
degrees of freedom of the trapped ions provides the initial conditions for the
chemical reactions. The neutral reagent gas is added to the trap as typically
a minor constituent to an inert buffer gas such as helium. Instead of giving a
full account of the recent research, the capabilities of multipole ion traps
are exemplified by discussing a few selected ion-molecule reaction
experiments.

\subsection{Reactions of hydrogen molecular ions}

The experimental procedure to study ion-molecule reactions is illustrated with
one of the conceptually simplest bimolecular ion-molecule reactions, the
hydrogen transfer reaction
\begin{equation}
{\rm H}_2^+ + {\rm H}_2 \rightarrow {\rm H}_3^+ + {\rm H}.
\label{h3formation:eq}
\end{equation}
With this exothermic reaction the triatomic hydrogen ion H$_3^+$ is formed in
hydrogen plasmas, one of the most important examples being cold molecular
clouds in the interstellar medium. This reaction is studied by trapping
H$_2^+$ ions in a multipole ion trap and adding a controlled density $n_{\rm
H_2}$ of H$_2$ through a leak valve to the buffer gas in the trap. The number
of H$_2^+$ ions will then decrease and H$_3^+$ ions will be formed as a
function of storage time. By extracting all ions from the trap after different
hold times and mass analyzing them one obtains data such as the one depicted
in the left panel of Fig.\ \ref{reactions:fig}. The observed exponential decay
of the number $N$ of H$_2^+$ ions follows reaction kinetics that are described
by a pseudo-first order rate equation
\begin{equation}
\frac{dN}{dt} = - k n_{{\rm H}_2} N,
\end{equation}
because the hydrogen density $n_{{\rm H}_2}$ stays practically constant during
the experiment. Here, $k$ is the reaction rate coefficient, the quantity of
interest that characterizes the collision process. For this reaction the rate
coefficient agrees well with the Langevin rate coefficient (see section
\ref{buffergas:sect}) \cite{glenewinkel97}. The probability that a reaction
occurs once the two collision partners have surmounted the centrifugal barrier
at long range is therefore near 100\%. This shows that there is no potential
barrier that could suppress this elementary chemical reaction.

The highly efficient formation of H$_3^+$ makes it the most important
molecular ion in the interstellar medium. In dense interstellar clouds it
drives the formation of larger molecules by proton transfer reactions
\cite{Smith1995Ions}. With the low temperatures of below 50\,K that prevail in
the dense interstellar clouds, isotopic exchange reactions of H$_3^+$ become
important. These reactions are driven by differences of the vibrational
zero-point energies of hydrogen and deuterium containing
molecules. Understanding these reaction rates and the influence of the
rotational quantum states, which are coupled by symmetry to the molecular
hyperfine states, on the reactivity still poses significant problems today
\cite{gerlich2002:pss,hugo2009:unp}.

\subsection{Negative ion-molecule reactions}

For negative ion-molecule reactions a very different situation is found
compared to the hydrogen molecular ion. A specific example is the proton
transfer reaction to the negatively charged amide ion
\begin{equation}
{\rm NH}_2^- + {\rm H}_2 \rightarrow {\rm NH}_3 + {\rm H}^-.
\label{nh2:reaction}
\end{equation}
As for most negative ion-molecule reactions
\cite{depuy2000:ijm,laehrdahl2002:ijms} this reaction is governed by a complex
Born-Oppenheimer potential energy hypersurface
\cite{kraka86,otto2008:prl}. Specifically, the reaction has to pass two deep
minima and an intermediate potential barrier on its way from reactants to
products. At room temperature a reaction probability, given by the ratio of
the measured rate coefficient to the calculated Langevin rate coefficient, of
about 2\% is found \cite{bohme1973:jcp}. Thus, 98\% of the collision events
lead to back-scattering of the reactants without further reaction. This occurs
despite the fact that the reaction is exothermic and also the intermediate
potential barrier lies below the energy of the entrance channel. It shows that
for this reaction the reaction dynamics at short range are very important.

\begin{figure}[tb]
\begin{center}
\includegraphics[width=0.49\columnwidth]{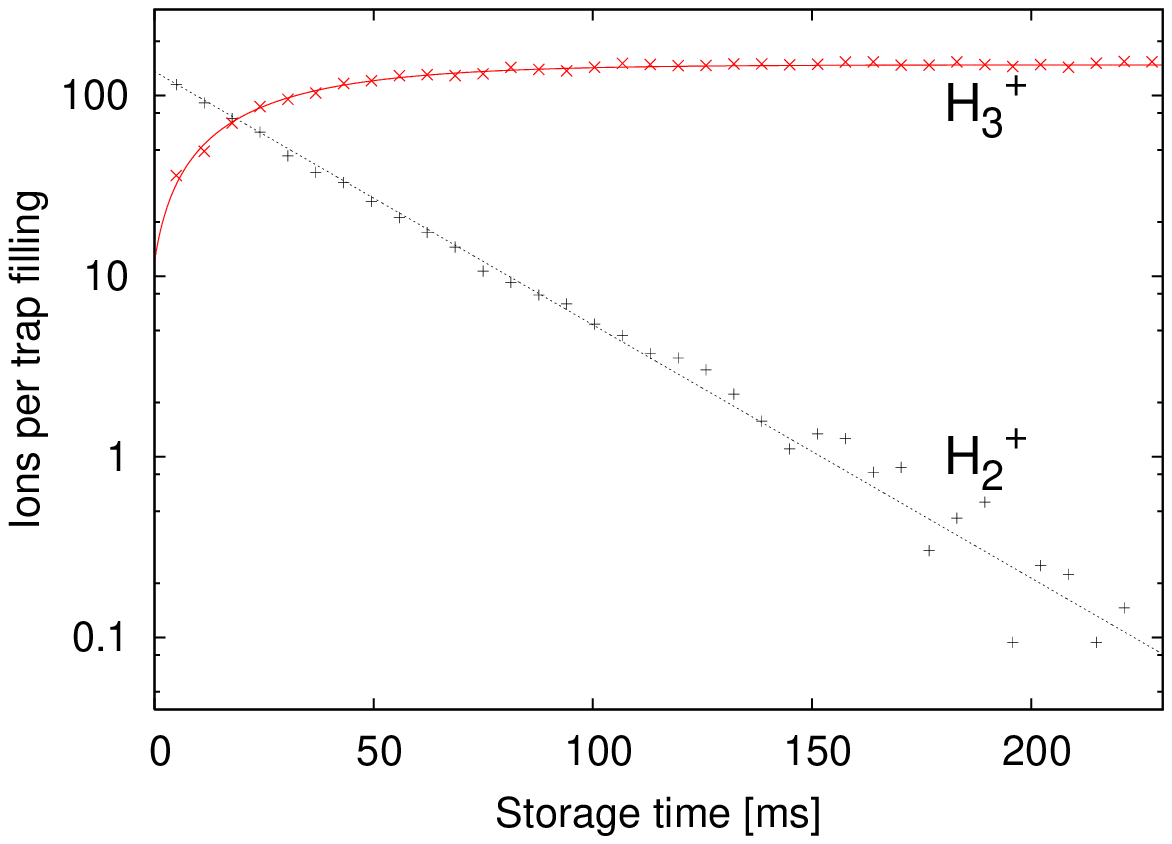}
\includegraphics[width=0.49\columnwidth]{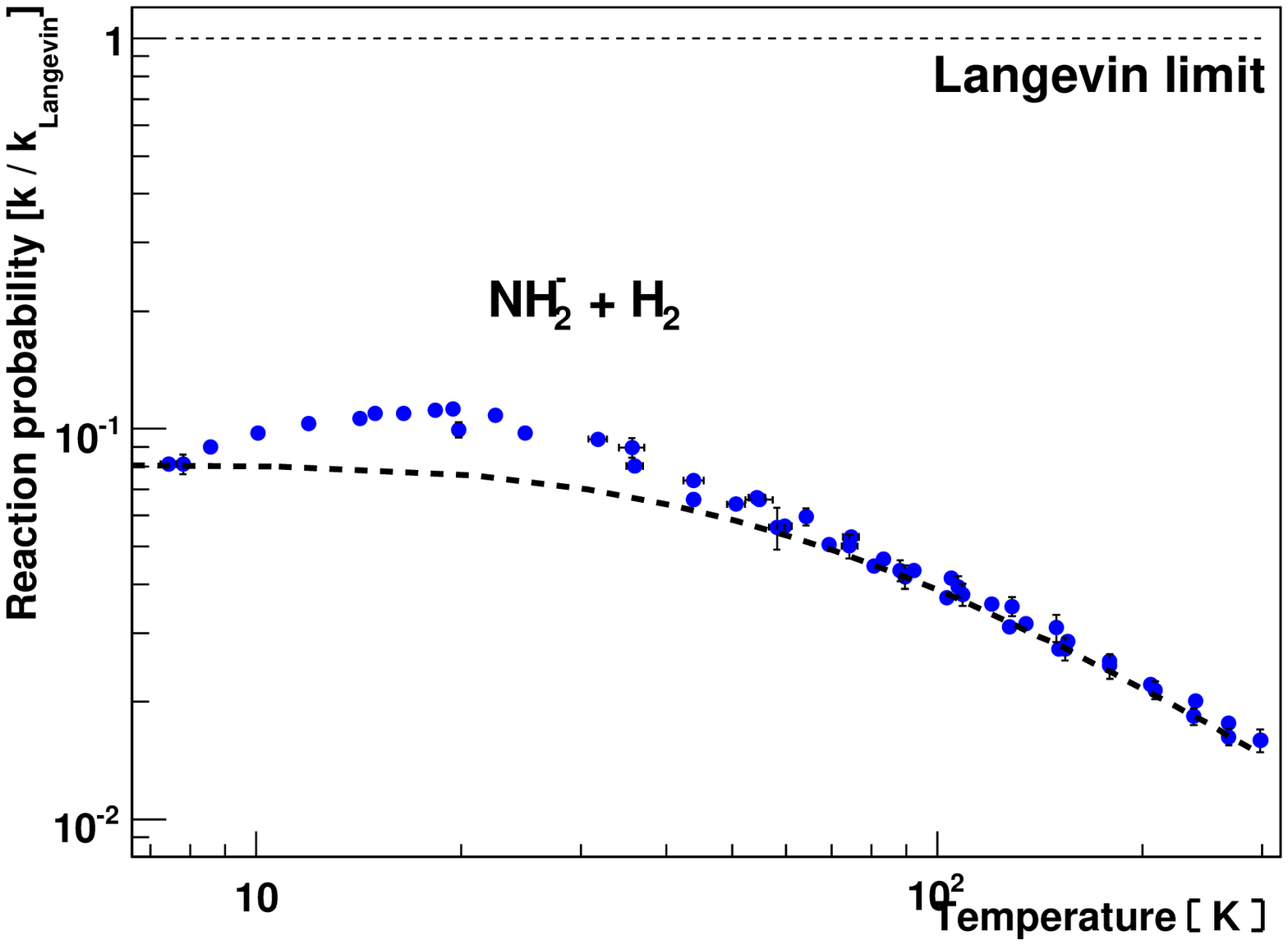}
\caption{\label{reactions:fig} (Color online) Left panel: Dataset of a typical
  ion-molecule reaction measurement in a 22-pole ion trap. The measured decay
  of H$_2^+$ due to reactions with neutral H$_2$ molecules
  (Eq. \ref{h3formation:eq}) and the corresponding formation of H$_3^+$ are
  shown as a function of storage time \cite{asvany2008:unp}. By fitting a rate
  coefficient model to the data (solid line) the reaction rate coefficient can
  be extracted. Right panel: Reaction probability, given by the ratio of the
  measured rate coefficient and the constant Langevin rate coefficient, for
  the reaction of NH$_2^-$ with H$_2$ (Eq. \ref{nh2:reaction}) as a function
  of temperature \cite{otto2008:prl}. The data show that the probability for
  reaction increases with decreasing temperature but stays well below the
  Langevin limit. Below 20\,K the data show an unexpected increase of the
  reaction probability that can not be explained by a classical statistical
  model (dashed line).}
\end{center}
\end{figure}

At lower temperatures the probability for reaction (\ref{nh2:reaction})
increases strongly, as measured in a 22-pole ion trap \cite{otto2008:prl}.
The data are shown in right panel of Fig.\ \ref{reactions:fig}. At 20\,K the
probability has increased by a factor of six. This increase is a manifestation
of the complex-mediated reaction dynamics \cite{troe1994:far}: the
intermediate NH$_4^-$ complex, which is transiently formed during a collision
that surmounts the centrifugal barrier, has a longer lifetime with respect to
decay back to reactants at lower temperatures, because the number of available
decay channels decreases. The probability to cross the intermediate potential
barrier and form products, however, remains approximately constant. Therefore
the overall probability to react increases.

The decrease of the reaction probability for temperatures lower than 20\,K can
not be explained within the classical dynamics picture of a complex-mediated
reaction mechanism. Instead it is expected to represent a signature of quantum
mechanical reaction dynamics in low temperature ion-molecule reactions
\cite{dashevskaya2005:jcp}.

\subsection{Complex formation in ternary collisions}

A process that connects molecular reaction dynamics with cluster physics is
the association of ionic complexes upon three-body collisions. Different types
of clusters have been studied in multipole ion traps, starting with
experiments on the formation of hydrogen clusters H$_5^+$ to H$_{23}^+$ at
10\,K \cite{paul1995:ijm}. Further ion trap experiments have investigated the
temperature-dependent growth of (CO)$_n^+$ clusters for $n=2$ to 9
\cite{schlemmer2002a:jcp} and the formation of Cl$^-\cdots$ CH$_3$Cl complexes
\cite{mikosch2008:jpc}.

The growth of ionic clusters in ternary collisions is modeled by two
subsequent reaction steps:
\begin{equation}
  \label{3body_equations}
  \begin{array}{c c c c c}
    & k_{f} n_{\rm B} & & k_{s} n_{\rm B} & \\
    \mathrm{A^+ + 2 \, B} & \stackrel{\rightharpoonup}{\leftharpoondown} &
    \mathrm{[A^+\cdots B]^* + B} & \rightarrow & \mathrm{[A^+\cdots B] + B}\\
    & \Gamma & & &\\
  \end{array}
\end{equation}
Under the conditions in an ion trap, low buffer gas densities and long
interaction times, the metastable intermediate complex is much more likely to
decay back to reactants than to undergo a stabilizing collision. In this
well-defined low-pressure regime the total formation rate $R$ is given by
\begin{equation}
R = k_3 n_{\rm B}^2,
\end{equation}
where $k_3$ represents the ternary rate coefficient. Typical values range
between $10^{-26}$ and $10^{-30}$\,cm$^6$/s. The quadratic dependence on the
neutral density is an important experimental signature of ternary collisions.

The temperature dependence of $k_3$ provides insight into the coupling of the
molecular degrees of freedom during the collisions. For many ternary
collisions the temperature dependence agrees with predictions from statistical
models \cite{viggiano86}. This was also observed for the formation of trapped
(CO)$_2^+$ \cite{schlemmer2002a:jcp}. For the growth of (CO)$_3^+$, however,
the next step in the formation of carbon monoxide clusters, deviations from
the statistical model occur \cite{schlemmer2002a:jcp}. Also in the ternary
association of chlorine anions with CH$_3$Cl, a large deviation was found in
the temperature dependence of $k_3$ \cite{mikosch2008:jpc}. This indicates
that the molecular degrees of freedom couple differently than assumed in the
statistical model. Quantum mechanical calculations may be required to clarify
this.

In the low-density regime the ternary rate coefficient can be written as
\begin{equation}
k_3 = \frac{k_f k_s}\Gamma = k_f k_s \tau.
\end{equation}
The formation rate $k_f$ of the metastable complex is given by the Langevin
collision rate for polarizable neutral species and by a modified capture rate
for polar molecules. The stabilization rate $k_s$ is the product of the
capture collision rate and an, often unknown, stabilization efficiency $\beta
\leq 1$. Thus, if $\beta$ is known or can at least be estimated, the lifetime
$\tau = \Gamma^{-1}$ of the metastable complex $\mathrm{[A^+\cdots B]^*}$
becomes experimentally accessible from measurements of the ternary rate
coefficient. For the carbon monoxide clusters, $\beta=1$ has been assumed and
lifetimes of the metastable complexes between 0.1 and 4\,ns have been deduced
\cite{schlemmer2002a:jcp} for a temperature of 80\,K. For the metastable
[Cl$^-\cdots$ CH$_3$Cl]$^*$ complexes, $\beta$ was experimentally inferred to
be near 1 for stabilizing collisions with CH$_3$Cl. This lead to complex
lifetimes that increased from 18\,ps at 220\,K to 42\,ps at 150\,K
\cite{mikosch2008:jpc}. These experiments show that studies of trapped ions
over many seconds allows one to explore molecular lifetimes that are 12 orders
of magnitude shorter.


\section{Laser spectroscopy of cold molecular ions \label{spectroscopy:sect}}

The preparation of molecular ions in well-defined quantum states of their
rotational and vibrational motion by buffer gas cooling in multipole ion traps
is very useful for precise spectroscopic studies (see e.g.  Ref.\
\cite{Schlemmer1999Laser,Schlemmer2002Laser,asmis2003:sci,mikosch2004:jcp,asvany2005:sci,boyarkin2006:jacs,dhzonson2006:ijm,Glosik06,janssens2006:prl}). However,
direct absorption or fluorescence spectroscopy can not be used at number
densities of $\sim 10^3$\,cm$^{-3}$ \cite{saykally1988:sci}. Also more
sensitive four-wave mixing or cavity ring-down techniques are not sensitive
enough to detect photoabsorption in samples of hundreds of ions. Instead, a
suitable action spectroscopy method is needed. Two schemes are most widely
used, spectroscopy by chemical probing and resonant multiphoton fragmentation
spectroscopy.

Action spectroscopy by chemical probing uses the fact that molecular ions are
confined in the trap irrespective of their internal excitation. The internally
excited molecular ion can then undergo a chemical reaction within the time set
by the lifetime of the excited state, either due to radiative emission or
inelastic quenching with non-reactive buffer gas. The detection of the
reaction product is the signature of the absorption step. This is a sensitive
spectroscopic technique, because product ions can be measured with
single-particle detection efficiency. Even if no reaction occurred for a given
excitation step, buffer gas cooling back to the initial internal state
distribution offers subsequent chances for successful probing.

\begin{figure}[tb]
\begin{center}
\includegraphics[width=0.4\columnwidth]{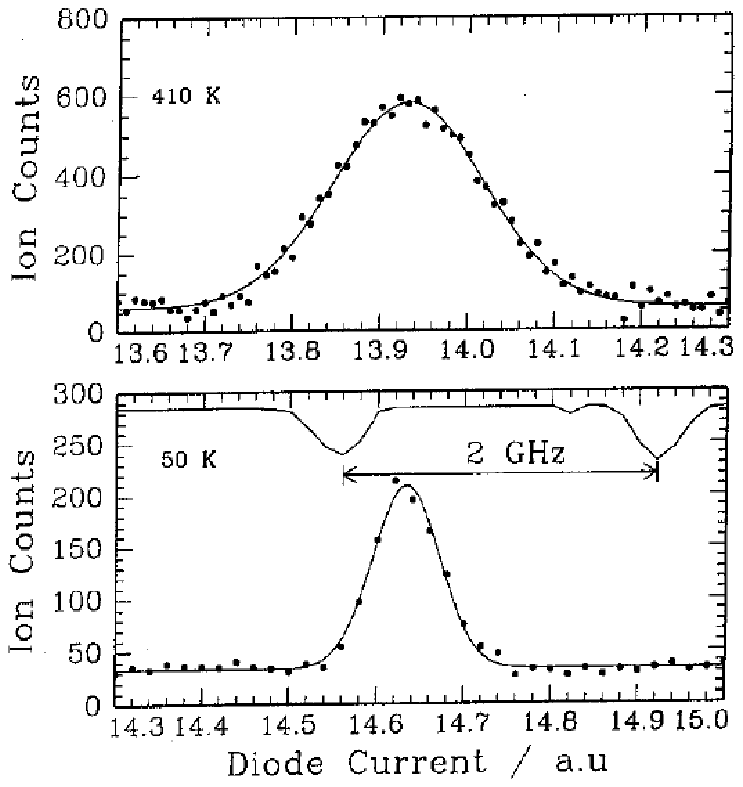}
\includegraphics[width=0.59\columnwidth]{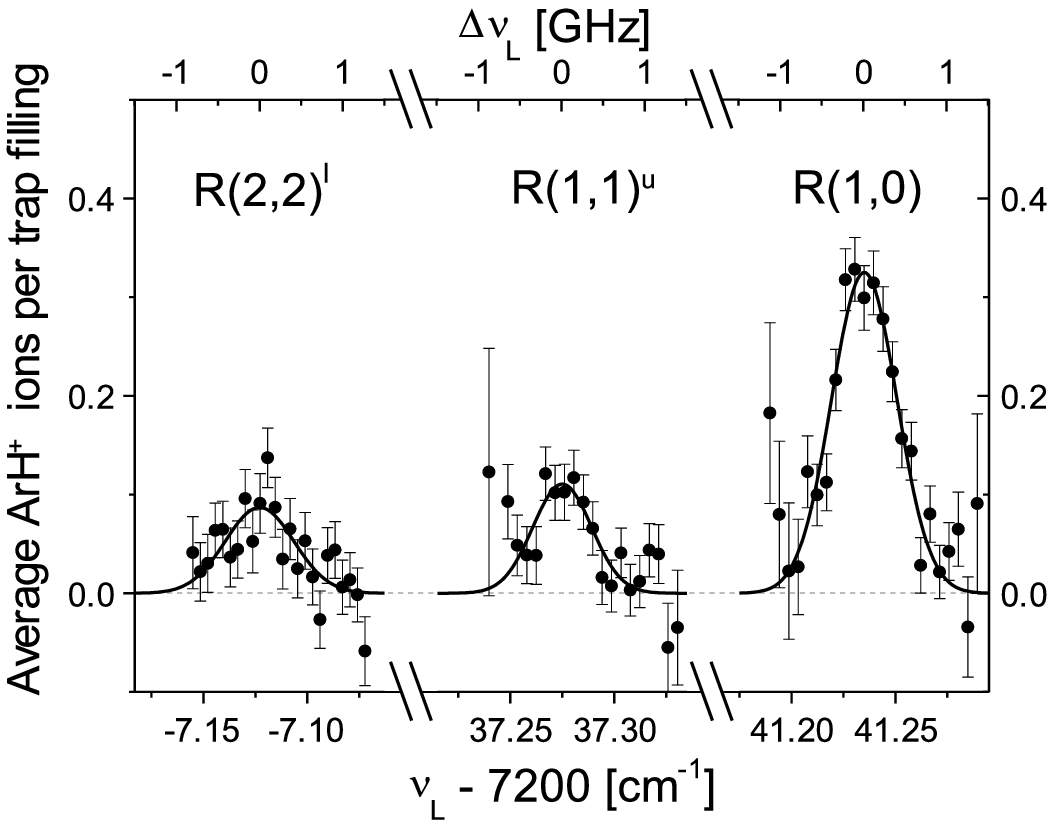}
\caption{\label{fig:spectra} Left panel: Electronic chemical probing
  spectroscopy of N$_2^+$. The spectroscopic signature is derived by
  laser-induced charge transfer in collisions with argon, forming Ar$^+$. Two
  line scans are shown with the Ar$^+$ count rate plotted as a function of
  laser frequency. By fitting Doppler profiles to the lines translational
  temperatures are obtained that agree with the applied buffer gas temperature
  \cite{Schlemmer1999Laser} Right panel: Action spectroscopy of vibrational
  overtones of H$_3^+$, probed by proton-transfer reactions with argon atoms,
  which become exothermal for three quanta of vibrational
  excitation. Doppler-broadened lines for three different rovibrational
  transitions are shown \cite{mikosch2004:jcp}. Here, the rotational and
  translational temperatures agreed with each other, but were much larger than
  the buffer gas temperature, which is explained by strong rf heating
  \cite{asvany2009:ijm}.}
\end{center}
\end{figure}

The earliest applications of chemical probing as a sensitive technique for
action spectroscopy employed ion beams \cite{carrington1978:mp,wing1976:prl}
and ion flow tubes \cite{grieman1982:cpl}. It has been applied for the first
time in a multipole ion trap to electronic spectroscopy of N$_2^+$ ions using
narrow-band continuous wave laser light \cite{Schlemmer1999Laser}. Here, the
endothermic charge transfer reaction of N$_2^+$ with Ar does not occur if the
N$_2^+$ ions are in the electronic and vibrational ground state and have a
near-room temperature thermal population of rotation states. Once the N$_2^+$
ions are excited to the A$^2\Pi_u$ state and have decayed back to
vibrationally excited levels in the X$^2\Sigma_g^+$ state, the charge transfer
process occurs with a finite rate coefficient and Ar$^+$ ions are
formed. After a suitable interaction time of the excitation laser with the
ions, the trap is emptied and the number of produced Ar$^+$ ions is
measured. By scanning the laser frequency across the Doppler profile of the
trapped ions one obtains the resonance profiles shown in the left panel of
Fig.\ \ref{fig:spectra}. Both at 410 and at 50\,K the widths of the Doppler
profiles are found to be in agreement with the expected width based on the
denoted buffer gas temperature determined by the trap wall. This shows that in
this experiment no significant radiofrequency heating occurred within the
experimental accuracy.

An alternative chemical probing scheme has been applied to high resolution
infrared overtone spectroscopy of the triatomic H$_3^+$ ion, again using a
narrow-band cw laser. Electric dipole-allowed transitions in this simplest
polyatomic molecule are limited to excitations of its asymmetric stretching
mode. Once more than two quanta of vibration are excited, the H$_3^+$ ion may
undergo proton transfer to Ar. The spectroscopic signature is therefore the
detection of ArH$^+$ product ions. Doppler profiles for the three lowest
rotational energy levels of H$_3^+$ $(J,K)$ = $(1,1)$, $(1,0)$ and $(2,2)$ are
plotted in the right panel of Fig.\ \ref{fig:spectra}. These data show that
the ions' Doppler temperature (found here to be about $150\pm20$\,K) does not
automatically coincide with the buffer gas temperature (set to 55\,K in this
experiment). This can be explained by rf heating due to H$_3^+$-Ar collisions,
which have an unfavorable mass ratio (see section \ref{buffergas:sect})
\cite{asvany2009:ijm}. This is circumvented using higher helium buffer gas
densities, as shown in a recent experiment that studies high overtone
excitation of H$_3^+$ levels that lie beyond the barrier to linearity
\cite{kreckel2008:jcp}.

Many more molecules have been studied with infrared chemical probing
spectroscopy, such as deuterated variants of H$_3^+$
\cite{Glosik06,asvany2007:jcp}, C$_2$H$_2^+$ \cite{Schlemmer2002Laser} and
CH$_5^+$ \cite{asvany2005:sci}. In the experiment on CH$_5^+$, the use of the
infrared free electron laser FELIX has allowed for an investigation of the
very floppy low-frequency bending vibrations. Besides yielding structural
information these chemical probing experiments may also reveal quantum-state
specific rate coefficients for the reaction that is induced by the laser. This
has already been used in Ref.\ \cite{Schlemmer2002Laser} to study the
influence of the rotational quantum state on the reactivity.

The second scheme for action spectroscopy of trapped ions is resonant
multiphoton fragmentation spectroscopy. It is widely applied in the
near-visible range, typically using nanosecond pulsed lasers. It can be
applied to a large number of ions that have low-lying excited electronic
states, in particular many organic molecules. Here the energy of two or three
photons is sufficient to fragment the molecule. With the first photon a
specific resonance is excited in the molecular ion that is studied by
mass-spectrometrically detecting the photofragments that are created after the
absorption of the additional photons. This approach is leading to new insight
into large carbon containing molecules \cite{jochnowitz2008:annu}, which may
be responsible for some of the unidentified diffuse interstellar bands, a long
standing problem in laboratory astrophysics \cite{sarre2006:jms}. Furthermore,
rapid progress is being made in the understanding of the structure of
protonated amino acids, bare and micro-solvated with a few water molecules
\cite{boyarkin2006:jacs,mercier2006:jacs}. Spectroscopic experiments have also
been carried out on pure water clusters, where the spectroscopic signature of
fragmentation is already achieved by the absorption of a single infrared
photon \cite{wang2003:jpc}.

Multipole ion traps have also recently led to precise studies of the
photodetachment of trapped negative ions in a laser field
\cite{trippel2006:prl}. Here the absorption of a single visible or
near-ultraviolet photon induced the detachment process A$^-$ + h$\nu$
$\rightarrow$ A + e$^-$. This leads to a loss rate for the trapped anions. By
scanning the entire ion density distribution with the photodetachment laser
beam, the absolute total photodetachment cross section can be extracted from
the density-dependent loss rate (see Ref.\ \cite{otto2009:subm} for more
details). This allows one to derive absolute cross sections with high accuracy
independent of any theoretical model. About 5\% relative accuracy have been
achieved for the O$^-$ and OH$^-$ cross sections in this way
\cite{hlavenka2009:acc}. These data are not only benchmarks for high-level
quantum calculations, they are also important parameters for modeling the
negative ion composition in the Earth's atmosphere or the stability of
interstellar negative ions \cite{mccarthy2006:apj,bruenken2007:apl}.


\section{Outlook \label{outlook:sect}}

Multipole ion traps are versatile tools to study different molecular
processes, in particular the 22-pole ion trap is being used in many
experiments. Recently a lot of progress is made in the understanding of the
dynamics of ions in multipole ion traps. Cooling and radiofrequency heating of
trapped ions in the buffer gas has recently been studied
\cite{asvany2009:ijm}. The density distribution of trapped ions and thus the
effective potential of multipole traps can now be visualized using
photodetachment tomography \cite{otto2009:subm}. With this knowledge new,
improved designs of ion traps may be anticipated. 

\begin{figure}[tb]
\begin{center}
\includegraphics[width=0.42\columnwidth]{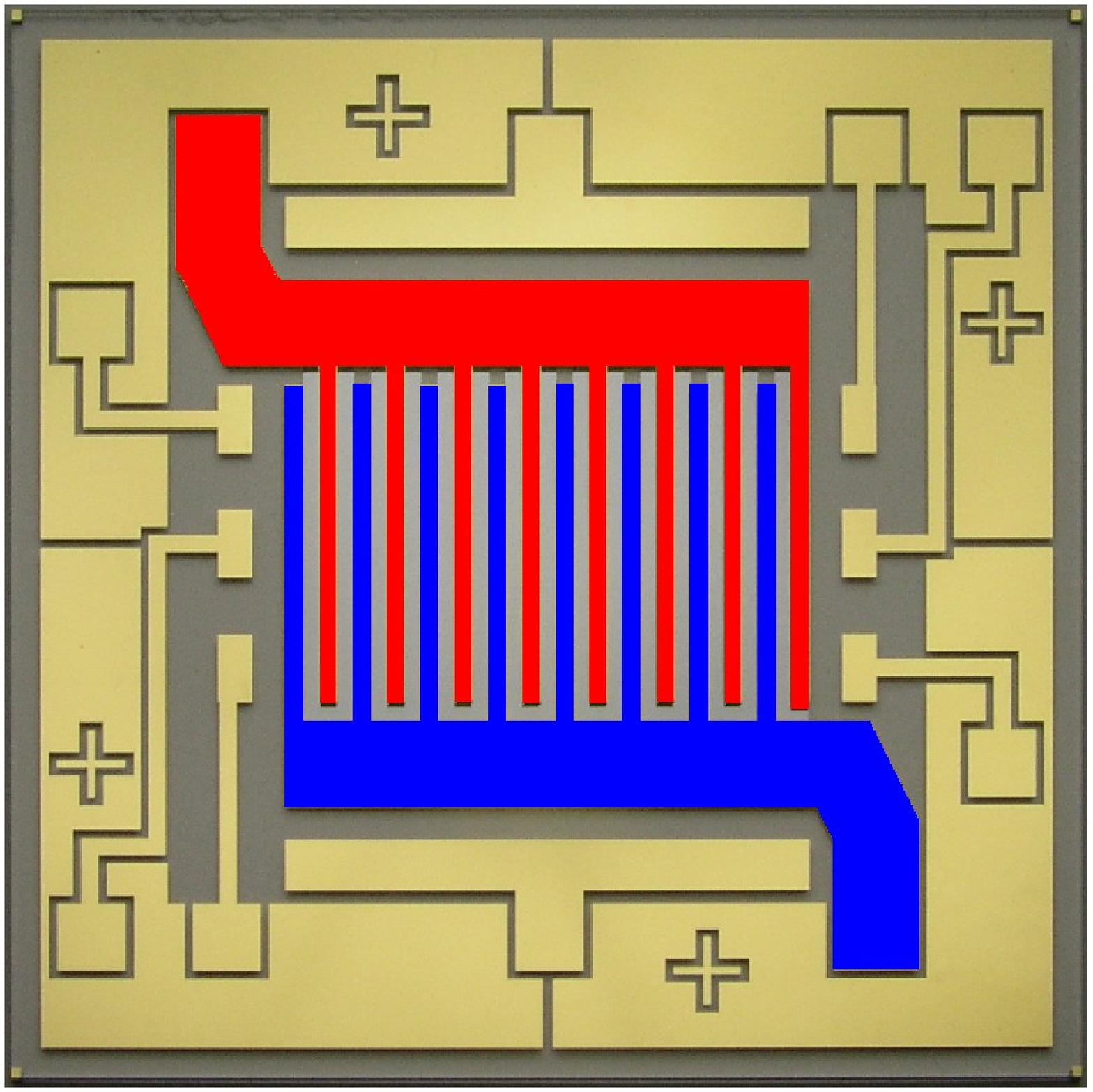}
\hspace{2mm}
\includegraphics[width=0.49\columnwidth]{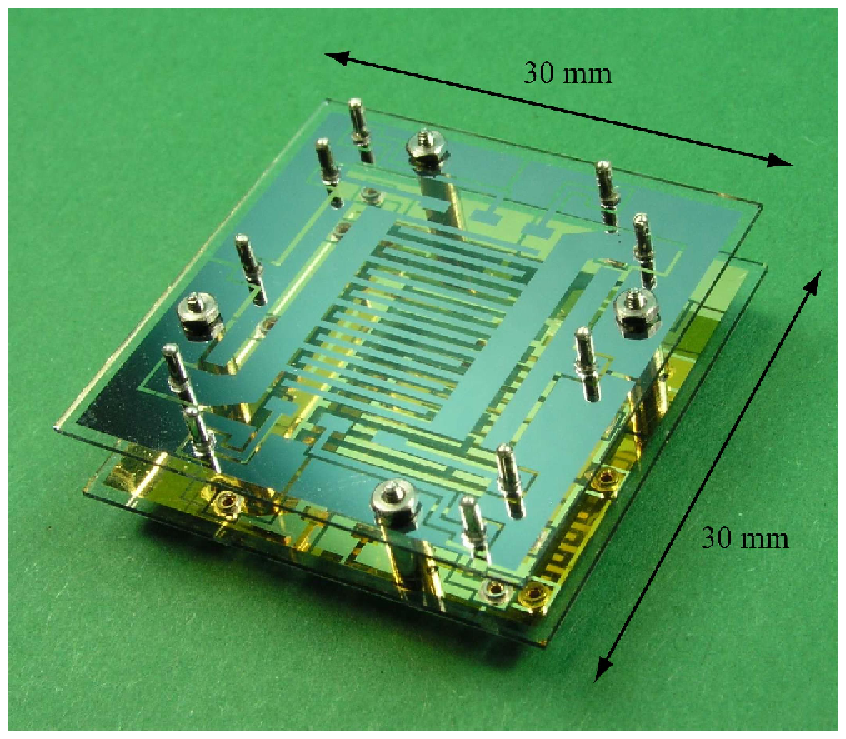}
\caption{\label{fig-planar-trap} (Color online) Left panel: view of one of the
  chips of a planar multipole ion trap \cite{kroener2007:td}. The gold
  electrodes that are connected to the two phases of the confining
  radiofrequency are colored in red and blue; they provide a repulsive
  effective potential ``mirror'' in front of the chip surface. Static
  potentials are applied to the additional electrodes on the chip in order to
  confine ions in all three dimensions. Right panel: photograph of an
  assembled chip-based planar multipole ion trap fabricated with $2\times16$
  etched gold electrodes on two glass substrates, 5\,mm apart. Ions are
  confined in the space between the two chips \cite{debatin2008:pra}.}
\end{center}
\end{figure}

New technologies for the fabrication of radiofrequency ion traps on planar
chip structures are already being explored extensively
\cite{madsen2004:apb,blain2004:ijm,stick2006:nat,pearson2006:pra,schulz2006:fp,seidelin2006:prl,pau2006:prl,amini2008:arxiv}. We
have recently developed a high-order multipole ion trap which is based on two
planar microstructured chips (see Fig.\ \ref{fig-planar-trap})
\cite{kroener2007:td} and which successfully traps ions for many seconds
\cite{debatin2008:pra}. Here 32 radiofrequency and several electrostatic
electrodes are etched out of a gold layer on top of the two glass
substrates. The goal of lower ion temperatures may be reached with colder
helium cryostats or possibly with non-thermal buffer gas from the tail of an
effusive beam. Laser-cooled atoms could in the future also be used to
collisionally cool ions to much lower temperatures than the current limit of
around 4\,K.

Future experiments with the cold ions in multipole ion traps may search for
quantum scattering or Feshbach resonances and their possible control with
external fields \cite{bodo2002:prl,stoecklin2005:pra,cote2000:pra}. They may
resolve the details of the rotational and hyperfine couplings in isotopic
substitution reactions of ``simple'' hydrogen molecular ions. And they may
cover processes that have recently been identified as important in
interstellar space, such as reactions of atomic species such as hydrogen,
carbon or oxygen or reactions of negatively charged molecular ions
\cite{millar2007:apj}.

The translationally and internally cold molecular ions that are prepared in a
multipole ion trap are also ideal for experiments that are carried out with
other experimental techniques. 22-pole ion traps are already employed as a
source at the electrostatic storage ring ELISA and the magnetic heavy ion
storage ring TSR \cite{kreckel2005:prl}. They also allow to prepare cold metal
clusters for photoelectron imaging studies \cite{bartels2009:sci}. In the
future they may be used for low-energy collision experiments of molecular ions
and clusters in a crossed-beam imaging spectrometer
\cite{mikosch2006:pccp,mikosch2008:sci}.

The spectroscopy with cold trapped molecular ions will expand with respect to
the systems studied and with respect to the employed frequency range of the
exciting radiation. The application of infrared and optical spectroscopy will
continue to give insight into vibrational and electronic transitions of linear
and cyclic organic molecules and in protonated polypeptides. In a pioneering
experiment pure rotational transitions excited with terahertz radiation have
been studied in trapped H$_2$D$^+$ \cite{asvany2008:prl}. In addition, new
ultraviolet and x-ray sources, such as the free-electron lasers, may allow for
new types of experiments with cold, trapped molecular ions.


\ack

This article reviews work that the author has been fortunate to carry out with
many colleagues. In particular collaboration with Jochen Mikosch, Sebastian
Trippel, Rico Otto, Petr Hlavenka, Ulrike Fr{\"u}hling, Raphael Berhane,
Christoph Eichhorn, Dirk Schwalm, and Matthias Weidem{\"u}ller is gratefully
acknowledged. We thank Dieter Gerlich for his advice on 22-pole traps and
Oskar Asvany and Stephan Schlemmer for many discussions on ion trapping. Our
research has been supported by the Elitef{\"o}rderung der Landesstiftung
Baden-W{\"u}rttemberg and by the Deutsche Forschungsgemeinschaft.


\section*{References}

\bibliographystyle{iopart-num}

\providecommand{\newblock}{}

\end{document}